\documentclass[]{article}
\pdfoutput=1
\usepackage[utf8]{inputenc}
\usepackage{jheppub}
\usepackage{graphicx}
\usepackage{amssymb}
\usepackage{amsmath}
\usepackage{bm}
\usepackage{color}
\usepackage{enumitem}
\usepackage{tensor}

\DeclareMathAlphabet\mathbfcal{OMS}{cmsy}{b}{n}
\definecolor{darkgreen}{RGB}{50,150,0}
\definecolor{purple}{cmyk}{0.5,0.75,0,0}
\definecolor{darkpurple}{RGB}{128,0,128}

\newcommand{\Mpl}{M_{\rm Pl}}

\definecolor{ultramarine}{rgb}{0.07, 0.04, 0.56}
\definecolor{cadmiumgreen}{rgb}{0.0, 0.42, 0.24}
\definecolor{indigo(dye)}{rgb}{0.0, 0.25, 0.42}
\usepackage[linktocpage=true]{hyperref}
\hypersetup{
colorlinks=true,
citecolor=ultramarine,
linkcolor=cadmiumgreen,
urlcolor=indigo(dye),
pdfauthor={},
pdftitle={},
pdfsubject={}
}

\title{Dark Energy and the Refined de Sitter Conjecture}

\begin{document}
\abstract{
  We revisit the phenomenology of quintessence
  models in light of the recently refined version of the de Sitter
  Swampland conjecture, which includes the possibility of unstable de
  Sitter critical points. We show that models of quintessence can
  evade previously derived lower bounds on $(1+w)$, albeit
  with very finely-tuned initial
  conditions. In the absence of such tuning or other rolling quintessence
  fields, a field with mass close to Hubble is required, which
  has a generic prediction for $(1+w)$. Slow-roll
  single field inflation models remain in tension. Other phenomenological
  constraints arising from the coupling of the quintessence field
  with the Higgs, pion or the QCD axion are significantly relaxed.
  \\
}
\author{Prateek Agrawal}
\author{and Georges Obied}

\affiliation{Jefferson Physical Laboratory, Harvard University
17 Oxford Street, Cambridge, MA 02138, USA}

\emailAdd{prateekagrawal@fas.harvard.edu}
\emailAdd{gobied@g.harvard.edu}

\maketitle

\section{Introduction}
A class of effective field theories (EFTs),
while otherwise consistent, do not admit a UV-completion within a
theory of quantum gravity.  Such EFTs are said to lie in the
Swampland~\cite{Vafa:2005ui}.  Delineating the boundaries of the
Swampland is an important task that could point to observable
consequences of a theory of quantum gravity.  Progress has been made
by studying the general properties of string compactifications (e.g. see
\cite{Banks:2010zn,Ooguri:2006in,ArkaniHamed:2006dz,
Ooguri:2016pdq,Obied:2018sgi} or~\cite{Brennan:2017rbf} for a recent
review).

One recent criterion is the de Sitter conjecture (dSC)
which proposes that all scalar potentials must satisfy the condition:
\begin{align}\label{eq:dSC}
  |\nabla V| \geq c\cdot V,
\end{align}
where the dimension-dependent constant $c\sim
\mathcal{O}(1)$. This condition is motivated by the
difficulty of de-Sitter (dS) constructions within string theory and
the plethora of no-go theorems which take the form of~\eqref{eq:dSC}
and thus forbid dS spacetimes under restricted
circumstances~\cite{Maldacena:2000mw,Hertzberg:2007wc,Wrase:2010ew,Andriot:2018ept}
(see also~\cite{Obied:2018sgi} and references therein).
However, for a different point of view, see for
example~\cite{Silverstein:2001xn,Maloney:2002rr,Kachru:2003aw,Cicoli:2018kdo,Akrami:2018ylq,Kachru:2018aqn,Hebecker:2018vxz,Conlon:2018eyr,Murayama:2018lie,Blaback:2018hdo}.

The dSC has observable consequences for dark energy\footnote{For a review of quintessence dark energy, see~\cite{Tsujikawa:2013fta}} equation of
state~\cite{Agrawal:2018own} and constrains single field slow-roll
inflation. This
has led to a number of model-building proposals~\cite{Ashoorioon:2018sqb,
Motaharfar:2018zyb,Dimopoulos:2018upl,Lin:2018kjm,Das:2018hqy,
Brahma:2018hrd,Damian:2018tlf,Matsui:2018bsy,Kehagias:2018uem,
Achucarro:2018vey,Lehners:2018vgi,Lin:2018rnx,Park:2018fuj,Ben-Dayan:2018mhe} to make
inflation consistent with dSC. Similarly, interest in quasi-dS
spaces consistent with dSC has motivated the construction of
quintessence models~\cite{Chiang:2018jdg,Wang:2018kly,DAmico:2018mnx,Han:2018yrk,Cicoli:2018kdo,Olguin-Tejo:2018pfq}.
Beyond these cosmological implications, a strong consequence arises
from studying the symmetric point of the
Higgs potential~\cite{Denef:2018etk,Murayama:2018lie,Hamaguchi:2018vtv,Han:2018yrk}, forcing
the quintessence field to
couple with the Higgs boson. A similar conclusion can be derived for
the QCD axion~\cite{Murayama:2018lie} and the
pion~\cite{Choi:2018rze}.

Alternate formulations of the dSC have also been proposed
in~\cite{Dvali:2018fqu,Andriot:2018wzk,Denef:2018etk}.
We study a particular refinement of the dSC (henceforth referred to as RdSC)
which has been proposed in~\cite{Garg:2018reu,Ooguri:2018wrx}.
The refinement allows~\eqref{eq:dSC} to be violated if the second
derivative of the potential is sufficiently negative. Explicitly, the
refined de Sitter conjecture states
\begin{align}
  \label{eq:RdSC}
  |\nabla V| \geq c\,V \quad
  {or}
  \quad
  \min(\nabla_i \nabla_j V)
  \leq
  -c' V
\end{align}
where $c,c'$ are constants of $\mathcal{O}(1)$.

The RdSC is motivated by its connection to the distance
Swampland conjecture~\cite{Ooguri:2006in} through Bousso’s covariant
entropy bound \cite{Bousso:1999xy}.
The refined conjecture evades all the counter-examples to the
dSC~\cite{Andriot:2018wzk,Roupec:2018mbn,Garg:2018zdg,Conlon:2018eyr}
since they involve tachyonic dS critical
points~\cite{Caviezel:2009tu,Flauger:2008ad}. Similarly, it also
evades constraints arising from coupling of the quintessence field
with the Higgs field, the pion and the QCD
axion~\cite{Denef:2018etk,Choi:2018rze,Murayama:2018lie}.

The original dSC had a firm prediction for the equation of
state for the dark energy.
In this note we investigate the implications of the
refined de Sitter conjecture (RdSC) for dark energy phenomenology.
We show that an arbitrarily fine-tuned initial condition can satisfy
current and future constraints on $w(z)$ for any values of $c$ and
$c'$, but generic initial conditions retain a prediction of
deviation from $w=-1$.

\section{Elsewhere on the moduli space}

Dark energy observations allow a quintessence field in the current
universe with a potential
whose slope is of the order of the vacuum energy (in Planck units).
Interesting constraints can be derived from considering the potential
of the quintessence field along with other scalar fields like the
Higgs and the pion (and potentially the axion) away from our current
position on the moduli space.

If the quintessence does not couple to the Higgs,
then the dSC is badly violated at the
symmetric point of the Higgs potential~\cite{Denef:2018etk}.
A coupling of the Higgs with the quintessence, on the other hand, leads
to larger-than-observed deviations in fifth force experiments, except for
perhaps a very fine-tuned set up where this coupling vanishes around
the Higgs minimum~\cite{Denef:2018etk,Murayama:2018lie}.
More recently it was shown that even this possibility is under tension
from time-dependence of the proton-to-electron mass
ratio~\cite{Hamaguchi:2018vtv}.

\begin{figure}[ht]
  \centering
  \includegraphics{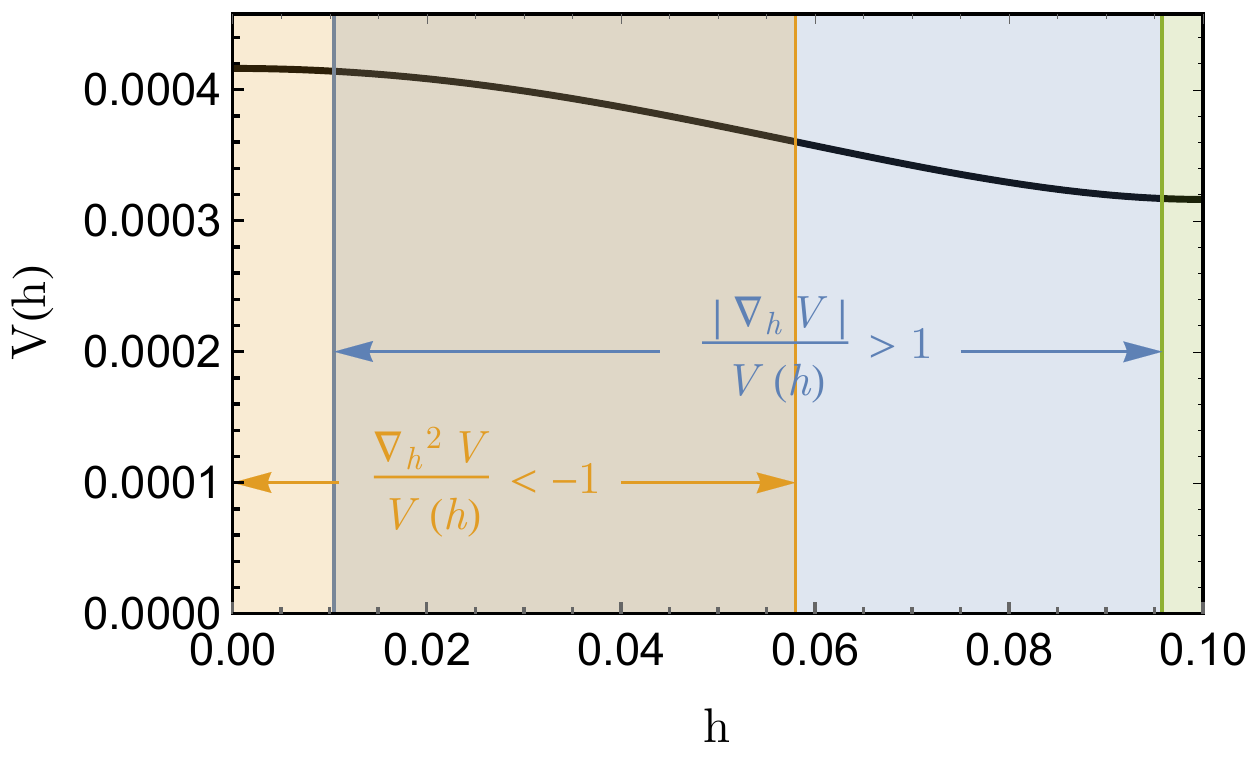}
  \caption{An example scalar
    potential (e.g.~ the Higgs potential) with spontaneous symmetry
    breaking, with parameters exaggerated for clarity. In the blue
    region, the (R)dSC is satisfied by the $h$ itself. In the orange
    region, the RdSC is satisfied along the $h$ direction. In the
    green region near the minimum along the $h$ direction, a
    ``quintessence'' field is needed to satisfy (R)dSC. If (R)dSC is
  satisfied at the minimum, then it is satisfied in all the green
region (see text for details).}
  \label{fig:higgspot}
\end{figure}

This tension is relaxed when we consider the refined conjecture.
Clearly, at the symmetric point on the Higgs potential the RdSC is
satisfied irrespective of the coupling of the Higgs with the
quintessence field. However, it is useful to see whether this
is true along the entire relevant range of the potential. For illustration
we take a toy potential,
\begin{align}
  V(\phi,h)
  &=
  \lambda_h(h^2-v^2)^2
  + V_0 e^{-\lambda \phi}
  \label{}
\end{align}
For values of $h$ outside a small
neighborhood of the origin (and away from the minimum), there is a slope in
the $h$ direction which satisfies the slope requirement in
(R)dSC,
\begin{align}
  \frac{\Mpl|\nabla_h V|}{V} > c
  \Rightarrow h \gtrsim \frac{c v^2}{4\Mpl},
\end{align}
where $c v^2/\Mpl \ll v$. However, near the origin, the second derivative
$\nabla_h^2 V$ causes the RdSC to be satisfied.
The latter switches sign at $h =
v/\sqrt{3}$. So the second derivative is sufficiently negative for
$h\sim v/\sqrt{3} \gg v^2 /\Mpl$,
\begin{align}
  \frac{\Mpl^2\nabla^2_h V}{V} < -c'
  \Rightarrow h \lesssim \frac{v}{\sqrt{3}}\left[1 - \frac{c'}{18}\left(\frac{v}{\Mpl}\right)^2 \right]
\end{align}
Finally, as the $h$ field nears its minimum at $v$, neither of the
above conditions is fulfilled, and the (R)dSC is satisfied by the
slope in the $\phi$ direction, as long as $\lambda > c$. The norm of
the gradient of
the potential is sufficiently large even away
from the minimum (and well into the region where $|\nabla_h V|$ is
large), if the following condition is met for the potential at $h=v$,
\begin{align}
  V(\phi, v) \lesssim \frac{8 \lambda_h}{\lambda^2} v^2 \Mpl^2
  \,.
\end{align}
This is a very mild restriction on the potential. 
This analysis can be carried out in essentially the same way for the SM
Higgs or the pion, as well as the QCD axion.
In figure~\ref{fig:higgspot} we show a toy example of this phenomenon.
We have chosen exaggerated numerical values for illustration. We see
that the RdSC is satisfied everywhere along the potential
with no cross coupling requirement between $\phi$ and $h$.

Finally, as noted in~\cite{Ooguri:2018wrx}, the RdSC also resolves
issues pointed out
by~\cite{Conlon:2018eyr} and, in a similar higher dimensional setting,
by~\cite{Murayama:2018lie}.
The general arguments in these papers only tell us about the presence
of an extremum, and in particular a maximum / inflection point.
Therefore, the RdSC can be easily consistent with these constructions
if the smallest eigenvalue of the Hessian is sufficiently negative at
these critical points.

\section{Dark Energy}

We now turn to an analysis of the implication of the refinement on
dark energy phenomenology. The dSC allows one to place a lower
bound on the
equation of state parameter $w\equiv p_\phi/\rho_\phi$ as presented
in~\cite{Agrawal:2018own}, and using current observations, an upper
bound on $c$.
We now show that the RdSC allows
 a class of models where
the quintessence field
satisfies the second clause of the refined conjecture,
and we cannot put a bound on $c$ using data.
Further, one can arrange for $w\approx -1$ today,
albeit at the expense of very finely
tuned initial conditions.
For concreteness we illustrate this
with an example potential for the quintessence field of the form,
\begin{align}
V(\phi)
&=
V_0
\left(a^4 - \frac{1}{2}b^2
\frac{\phi^2}{M_{\rm Pl}^2}\right)
\,,
\label{eq:qpot}
\end{align}
where $V_0 = (2.23\ {\rm meV})^4$ is the value of dark energy in
$\Lambda{\rm CDM}$.  This potential satisfies the second clause of the 
RdSC when $b^2 \geq
c' a^4$.  The mass of the
field can be conveniently written in terms of the Hubble parameter
today, $m_\phi^2 = -b^2 V_0/\Mpl^2 = -3 b^2 \Omega_\Lambda H_0^2$,
where $\Omega_\Lambda = 0.692$ is the dark energy fraction of the
universe today.
Given the dark energy content of the universe today, we see that this
potential requires $a\gtrsim 1$. Then RdSC puts a lower bound on the
curvature of
the quintessence potential, $b^2 > c'$,
or $|m_\phi^2| > 3c'\Omega_\Lambda H_0^2$.

\begin{figure}[t]
  \centering
  \includegraphics[width=0.65\textwidth]{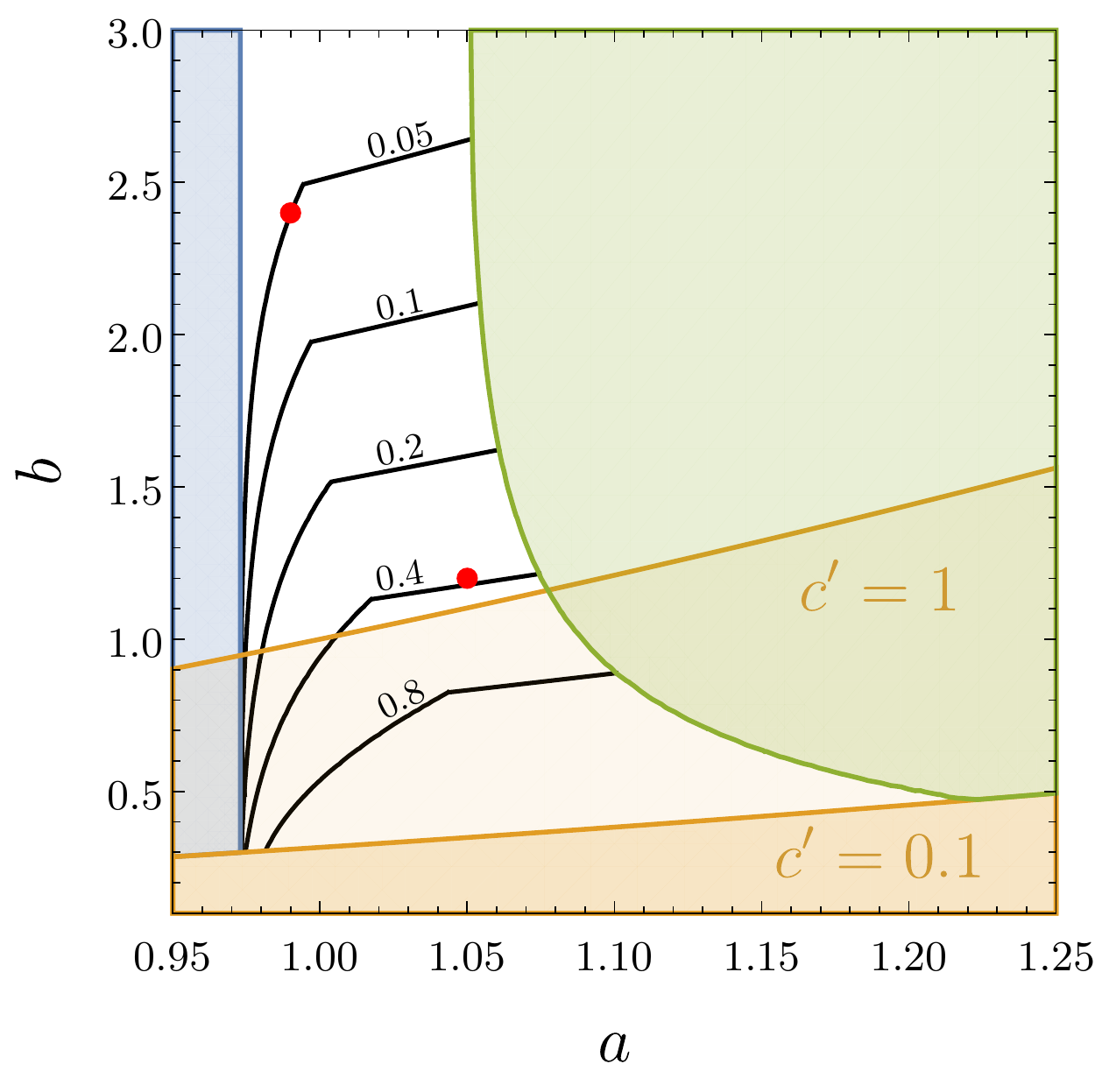}
\caption{Allowed part of parameter space for our
  potential~\ref{eq:qpot} subject to the RdSC (lower shaded), $w(z)$
  and $\Omega_\Lambda$ (upper
  right shaded) and $\Omega_\Lambda$ (left shaded)
  constraints.
  Contours show the maximum allowed misalignment of the field
  $\phi/\Mpl$ from the
  top of the potential. The contour segment to the left of the kink is
  constrained by $\Omega_\Lambda$, whereas the segment to the right is
  constrained by $w(z)$.
  Red dots indicate benchmark examples plotted in
  figures~\ref{fig:generic} and~\ref{fig:tuned}. We show regions
  excluded by the RdSC (orange) for $c' = 0.1$ and $c'=1$
  for illustration.
}
  \label{fig:abplot}
\end{figure}
If we ignore quantum fluctuations, then it is clear that we can set
the field at the top of the potential, and obtain $w = -1$ exactly.
However, quantum fluctuations will destabilize the field from the
maximum and fragment it to form domain walls. Therefore, we have to
ensure that the field is sufficiently misaligned from the top to
behave classically. The classical regime only requires a very tiny
misalignment,
\begin{align}
  \phi > H \frac{H^{2}}{3b^2\Omega_\Lambda H_0^2}.
\end{align}
At the same time, if the field has a large misalignment from the
maximum, its equation of state will deviate from $w = -1$.

There are two possible allowed regimes. The first is when $b\gg1$, the
slope of the
potential is large as soon as the field is misaligned from the
maximum, and leads to large deviations in $w+1$. In this case the
initial misalignment has to be very finely tuned to be consistent with
supernova observations. In the second case, near the boundary of the
RdSC constraint, $b^2 \gtrsim c' a^4$, the mass of the field is order
$H_0$, and therefore it only starts rolling today for $\mathcal{O}(1)$
misalignment, making it consistent with observations.

\begin{figure}[t]
  \centering
  \includegraphics[width=0.45\textwidth]{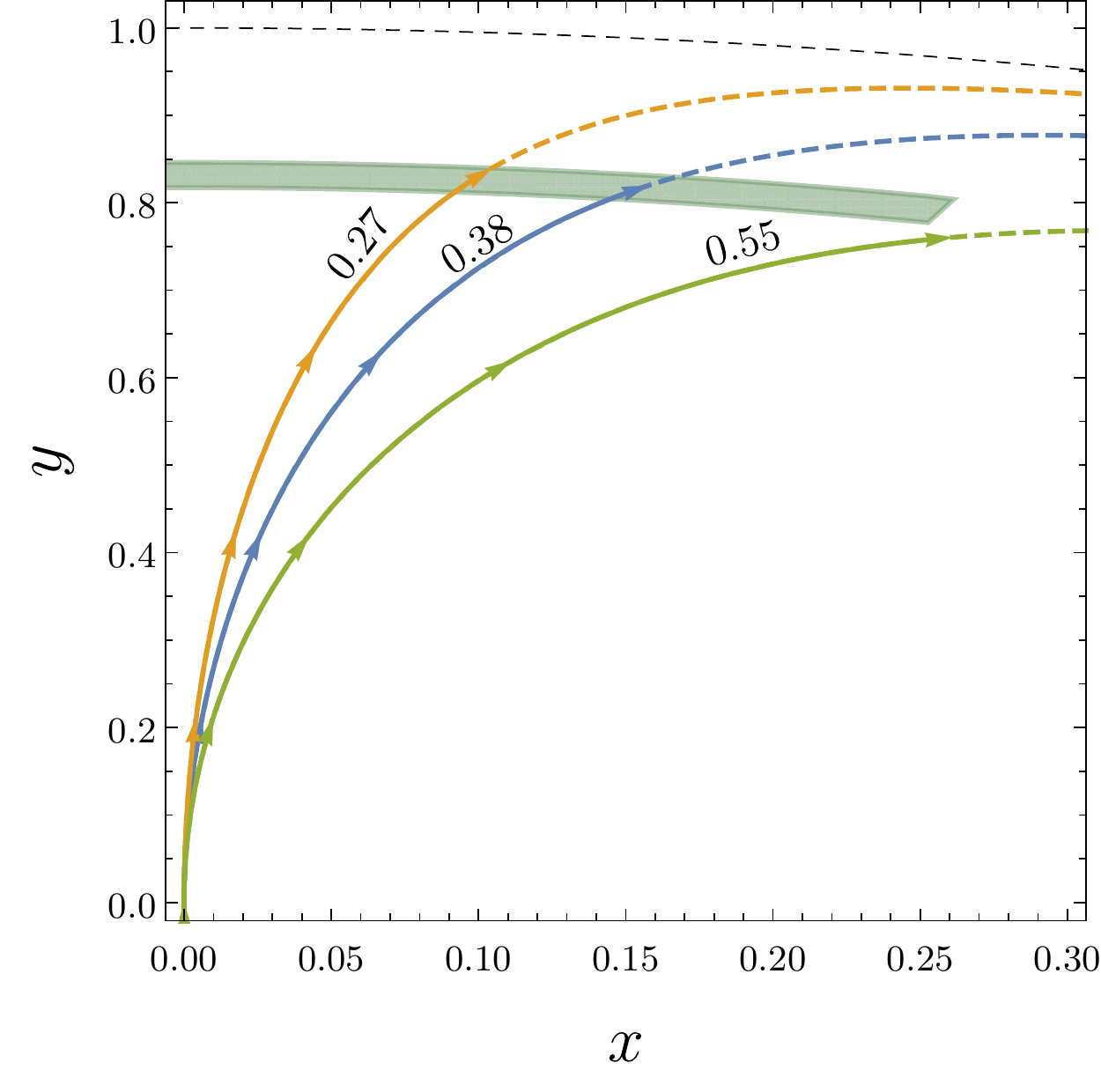}
  \qquad
  \includegraphics[width=0.475\textwidth]{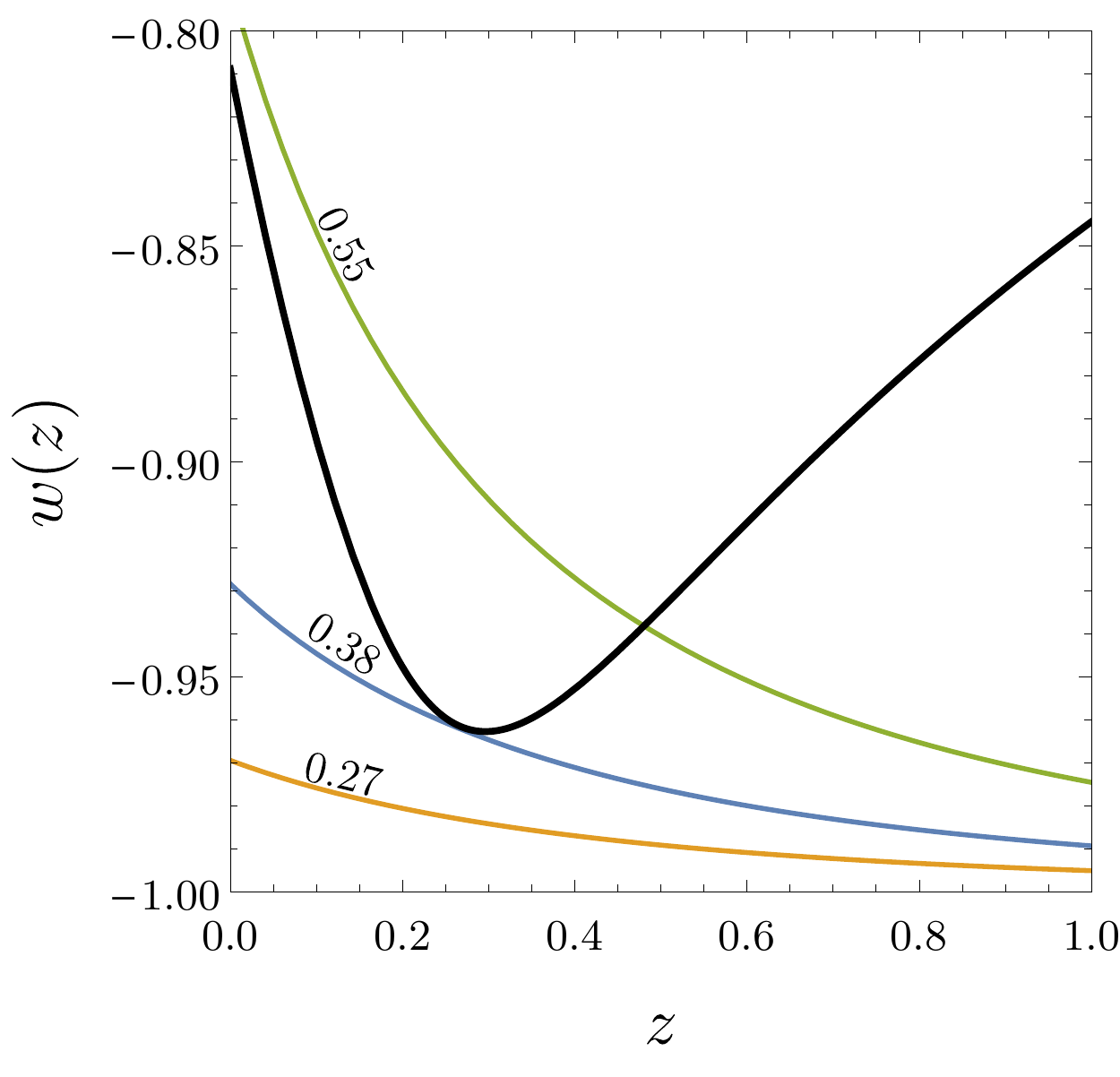}
  \caption{The $(x,y)$ trajectories of the quintessence field (left)
    and its equation of state (right) for a range
    of initial misalignment. For this plot, we have chosen the values
    $(a,b)=(1.05,1.2)$, for which generic initial conditions
    are consistent with data. Curves are labeled with the initial field value
    $\phi_{\rm init}/\Mpl$ and show the minimum and maximum allowed values
    in addition to a trajectory that is excluded by data.
  }
  \label{fig:generic}
\end{figure}

\begin{figure}[t]
  \centering
  \includegraphics[width=0.45\textwidth]{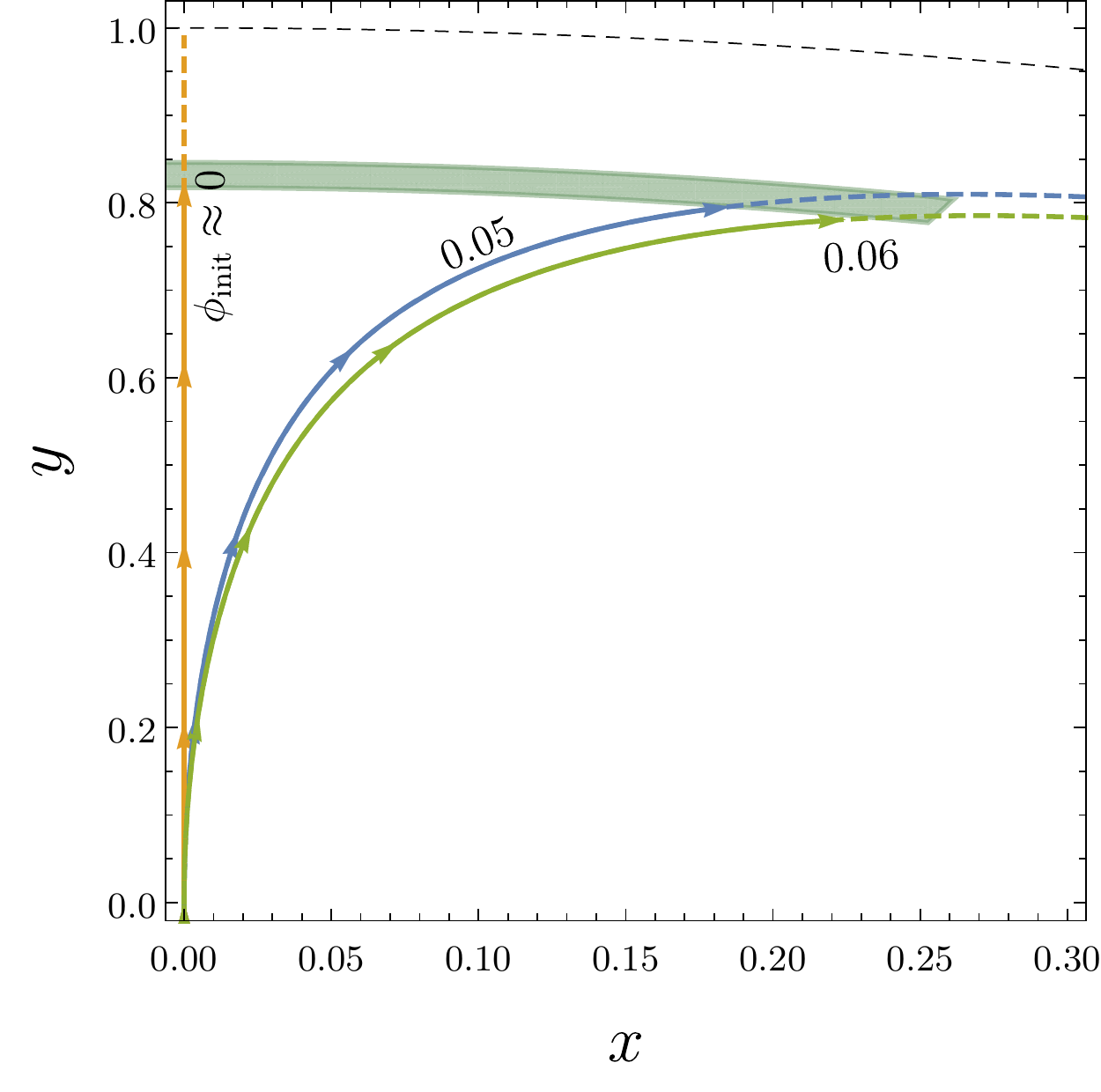}
  \qquad
  \includegraphics[width=0.475\textwidth]{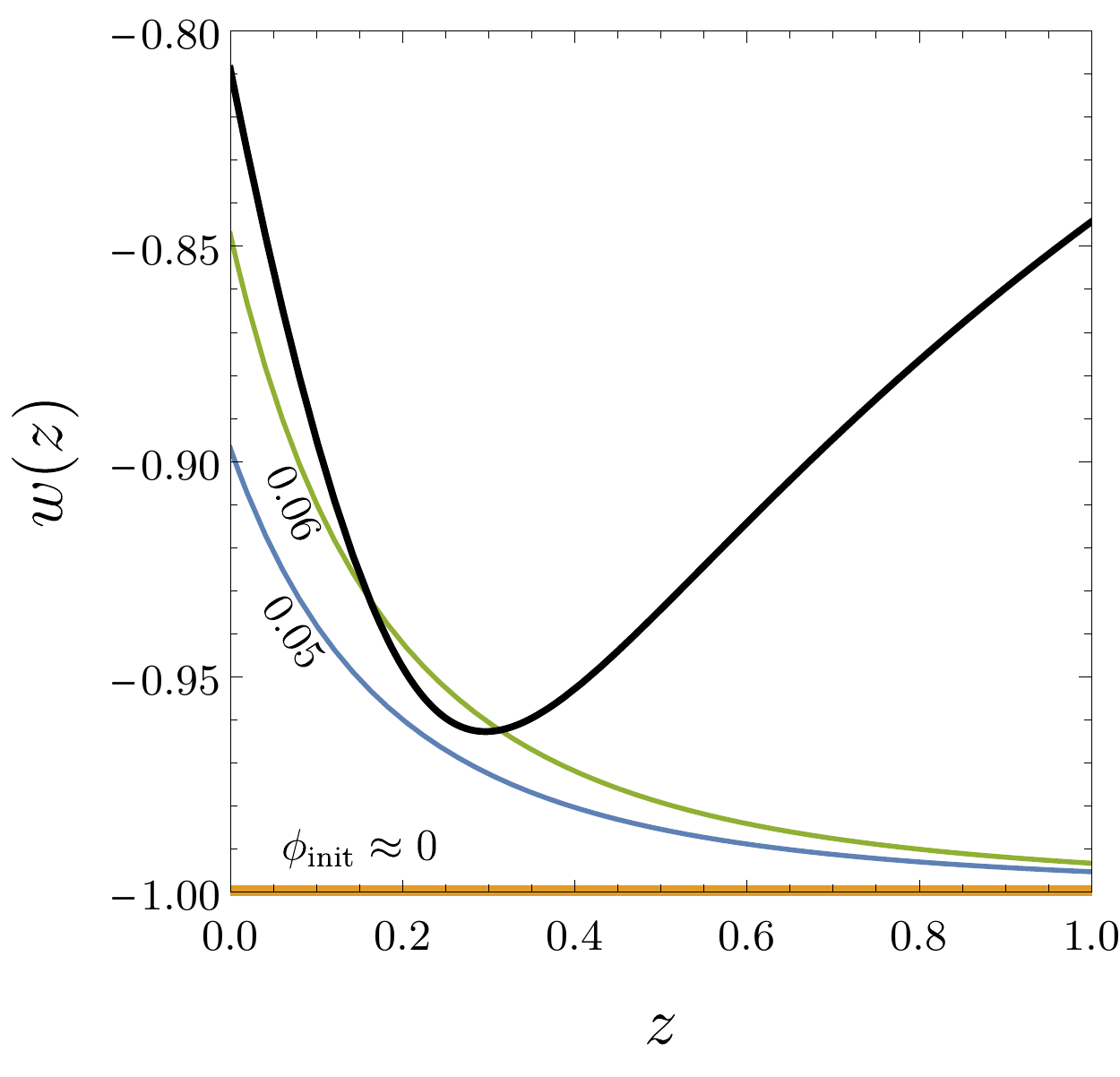}
  \caption{The $(x,y)$ trajectories of the quintessence field (left)
    and its equation of state (right) for a range
    of initial misalignment. For this plot, we have chosen the values
    $(a,b)=(0.99,2.4)$, for which only fine-tuned initial conditions
    are consistent with data. Curves are labeled with the initial field value
    $\phi_{\rm init}/\Mpl$ and show the minimum and maximum allowed values
    in addition to a trajectory that is excluded by data.
  }
  \label{fig:tuned}
\end{figure}

In figure~\ref{fig:abplot} we show the allowed parameter range in
$(a,b)$ space.
We also show the contours of the maximal initial value of the field
that is allowed by observations of
$\Omega_\Lambda$~\cite{Akrami:2018odb} and $w(z)$~\cite{Scolnic:2017caz}. As
in~\cite{Agrawal:2018own}, we choose the $2\sigma$ contours for $w(z)$
in~\cite{Scolnic:2017caz} and for $\Omega_\Lambda$
from \cite{Ade:2015xua}, and find the largest initial misalignment that is
consistent with the measurement of $\Omega_\Lambda$ and $w(z)$. For a
different procedure to extract $w(z)$ constraints to put a bound on
$c$, see~\cite{Heisenberg:2018yae, Heisenberg:2018rdu}.
For $b \gg 1$, we see that the initial values have to be tuned, as
noted above.  In the region $b\sim 1$, we can allow generic initial
conditions.
The trajectories of the fields for specific values of $(a,b)$ are
shown in
Figures~\ref{fig:generic} and~\ref{fig:tuned}.
We show the evolution of this system in $(x,y)$ coordinates defined as
\begin{align}
  x \equiv \frac{\dot{\phi}}{\sqrt{6}\Mpl H};
  \qquad y\equiv \frac{\sqrt{V}}{\sqrt{3}\Mpl H},
\end{align}
as well as the evolution of the equation of state $w$.
We have chosen  examples where the initial
conditions are allowed to be generic (Fig.~\ref{fig:generic}) and
where we need to tune the initial conditions to satisfy the supernova
constraints (Fig.~\ref{fig:tuned}).

We see that in either case, tuning the initial conditions allows us to
push $w$ as close to $-1$ as we want, evading the
lower bound derived in~\cite{Agrawal:2018own}.
However, in the absence of this initial condition tuning, a generic
prediction on the equation of state can be estimated. If $c' \gg 1$,
then it is generally hard to satisfy the current $w(z)$ constraints
with untuned initial conditions, and the least tuned initial
conditions will typically have $(1+w)$ close to the constraints today.
If $c'\lesssim 1$, then for $b^2 \sim c' a^4$ and $a \sim 1$ we can have
an $\mathcal{O}(\Mpl)$
misalignment. The slope of the potential in this case is given by
\begin{align}
  V'(\phi)
  &=
  V_0 b^2 \frac{\phi}{\Mpl}
  \approx \frac{V_0}{\Mpl} c'
\end{align}
Using the ``slow-roll'' approximation, $3H\dot\phi\simeq V'(\phi)$,
we can estimate the value of $w$ for this parameter
choice~\cite{Gott:2010xw,Slepian:2013ug},
\begin{align}
  1+w
  &\simeq
  \frac{\dot\phi^2}{V_0}
  \sim
  \frac{V'^2}{9 H^2 V_0}
  \sim
  \frac13\Omega_\Lambda c'^2
  \,.
\end{align}
It is interesting to compare this to a very similar looking bound
derived in~\cite{Agrawal:2018own}. We emphasize however that in the
current
case this is not a hard bound but more a generic prediction.
Due to this fact, we are unable to derive a robust bound on $c'$ using
current data as was done for $c$ in~\cite{Agrawal:2018own}. Further, we can trade off
bounds on $c$ with bounds on $c'$, and with tuned initial conditions
we can
have both $c$ and $c'$ to be $\mathcal{O}(1)$.

\section{Conclusion and discussion}
Swampland conjectures are in a very active phase of exploration. We
have studied the implications of the recently proposed refined de
Sitter conjecture~\cite{Ooguri:2018wrx,Garg:2018reu,Andriot:2018wzk,Dvali:2018fqu}.
The conjecture is motivated by an older distance Swampland conjecture
and its connection with Bousso's covariant entropy bound.
This conjecture circumvents a number of theoretical and phenomenological
tensions arising from coupling of the quintessence field to other
scalar fields in the standard model and beyond.
The RdSC is also consistent with constructions which are claimed to be
counter-examples to the earlier de Sitter conjecture.

As in dSC, the RdSC appears to be in tension with single-field
slow-roll inflation if both $c,c'$ are strictly $\mathcal{O}(1)$~\cite{Fukuda:2018haz}.
The fact that either $|\epsilon_V|$ or $|\eta_V|$ are $\mathcal{O}(1)$ makes it impossible
to naturally satisfy constraints on the scalar tilt $n_s -1 \approx
2\eta_V - 6\epsilon_V \approx 0$. In fact, recent
analyses~\cite{Akrami:2018odb} show that $\mathcal{O}(1)$ values for
either of $\eta_V$ or $\epsilon_V$ are strongly ruled out.
More extended
inflationary models can potentially evade this tension; these models
often come with detectable deviations from single-field case~\cite{Das:2018hqy,Das:2018rpg,Motaharfar:2018zyb}.
Unless the initial conditions are very fine-tuned, the
conclusions of~\cite{Agrawal:2018own} for the future cosmology of the
universe remain mostly unchanged.

It would be very interesting to try and identify models for such a
scalar field. In the string axiverse we expect a plenitude of light
scalars, and the positive dark energy of the universe in such a system
can be made up of a number of light particles.  What the RdSC adds to
this picture then is that it hints towards an axion of mass comparable
to the Hubble scale, with a misalignment from its minimum of
$\mathcal{O}(\Mpl)$ such that it has just recently begun dominating
the universe and appears briefly as dark energy, before eventually
starting to oscillate as matter.

\begin{acknowledgments}
We thank David Pinner for useful discussions. We are grateful to Paul
Steinhardt and Cumrun Vafa for guidance and useful comments on the
manuscript.
The work of PA is supported by the NSF grants
PHY-0855591 and PHY-1216270.
\end{acknowledgments}

\bibliographystyle{JHEP}
\bibliography{ref}

\providecommand{\href}[2]{#2}\begingroup\raggedright\begin{thebibliography}{10}

\bibitem{Vafa:2005ui}
C.~Vafa, \emph{{The String landscape and the swampland}},
  \href{https://arxiv.org/abs/hep-th/0509212}{{\ttfamily hep-th/0509212}}.

\bibitem{Banks:2010zn}
T.~Banks and N.~Seiberg, \emph{{Symmetries and Strings in Field Theory and
  Gravity}}, \href{https://doi.org/10.1103/PhysRevD.83.084019}{\emph{Phys.
  Rev.} {\bfseries D83} (2011) 084019}
  [\href{https://arxiv.org/abs/1011.5120}{{\ttfamily 1011.5120}}].

\bibitem{Ooguri:2006in}
H.~Ooguri and C.~Vafa, \emph{{On the Geometry of the String Landscape and the
  Swampland}},
  \href{https://doi.org/10.1016/j.nuclphysb.2006.10.033}{\emph{Nucl. Phys.}
  {\bfseries B766} (2007) 21}
  [\href{https://arxiv.org/abs/hep-th/0605264}{{\ttfamily hep-th/0605264}}].

\bibitem{ArkaniHamed:2006dz}
N.~Arkani-Hamed, L.~Motl, A.~Nicolis and C.~Vafa, \emph{{The String landscape,
  black holes and gravity as the weakest force}},
  \href{https://doi.org/10.1088/1126-6708/2007/06/060}{\emph{JHEP} {\bfseries
  06} (2007) 060} [\href{https://arxiv.org/abs/hep-th/0601001}{{\ttfamily
  hep-th/0601001}}].

\bibitem{Ooguri:2016pdq}
H.~Ooguri and C.~Vafa, \emph{{Non-supersymmetric AdS and the Swampland}},
  \href{https://doi.org/10.4310/ATMP.2017.v21.n7.a8}{\emph{Adv. Theor. Math.
  Phys.} {\bfseries 21} (2017) 1787}
  [\href{https://arxiv.org/abs/1610.01533}{{\ttfamily 1610.01533}}].

\bibitem{Obied:2018sgi}
G.~Obied, H.~Ooguri, L.~Spodyneiko and C.~Vafa, \emph{{De Sitter Space and the
  Swampland}},  \href{https://arxiv.org/abs/1806.08362}{{\ttfamily
  1806.08362}}.

\bibitem{Brennan:2017rbf}
T.~D. Brennan, F.~Carta and C.~Vafa, \emph{{The String Landscape, the
  Swampland, and the Missing Corner}},
  \href{https://doi.org/10.22323/1.305.0015}{\emph{PoS} {\bfseries TASI2017}
  (2017) 015} [\href{https://arxiv.org/abs/1711.00864}{{\ttfamily
  1711.00864}}].

\bibitem{Maldacena:2000mw}
J.~M. Maldacena and C.~Nunez, \emph{{Supergravity description of field theories
  on curved manifolds and a no go theorem}},
  \href{https://doi.org/10.1142/S0217751X01003935,
  10.1142/S0217751X01003937}{\emph{Int. J. Mod. Phys.} {\bfseries A16} (2001)
  822} [\href{https://arxiv.org/abs/hep-th/0007018}{{\ttfamily
  hep-th/0007018}}].

\bibitem{Hertzberg:2007wc}
M.~P. Hertzberg, S.~Kachru, W.~Taylor and M.~Tegmark, \emph{{Inflationary
  Constraints on Type IIA String Theory}},
  \href{https://doi.org/10.1088/1126-6708/2007/12/095}{\emph{JHEP} {\bfseries
  12} (2007) 095} [\href{https://arxiv.org/abs/0711.2512}{{\ttfamily
  0711.2512}}].

\bibitem{Wrase:2010ew}
T.~Wrase and M.~Zagermann, \emph{{On Classical de Sitter Vacua in String
  Theory}}, \href{https://doi.org/10.1002/prop.201000053}{\emph{Fortsch. Phys.}
  {\bfseries 58} (2010) 906} [\href{https://arxiv.org/abs/1003.0029}{{\ttfamily
  1003.0029}}].

\bibitem{Andriot:2018ept}
D.~Andriot, \emph{{New constraints on classical de Sitter: flirting with the
  swampland}},  \href{https://arxiv.org/abs/1807.09698}{{\ttfamily
  1807.09698}}.

\bibitem{Silverstein:2001xn}
E.~Silverstein, \emph{{(A)dS backgrounds from asymmetric orientifolds}},  in
  \emph{{Strings 2001: International Conference Mumbai, India, January 5-10,
  2001}}, 2001, \href{https://arxiv.org/abs/hep-th/0106209}{{\ttfamily
  hep-th/0106209}},
  \href{http://www-public.slac.stanford.edu/sciDoc/docMeta.aspx?slacPubNumber=SLAC-PUB-8869}{http://www-public.slac.stanford.edu/sciDoc/docMeta.aspx?slacPubNumber=SLAC-PUB-8869}.

\bibitem{Maloney:2002rr}
A.~Maloney, E.~Silverstein and A.~Strominger, \emph{{De Sitter space in
  noncritical string theory}},  in \emph{{The future of theoretical physics and
  cosmology: Celebrating Stephen Hawking's 60th birthday. Proceedings, Workshop
  and Symposium, Cambridge, UK, January 7-10, 2002}}, pp.~570--591, 2002,
  \href{https://arxiv.org/abs/hep-th/0205316}{{\ttfamily hep-th/0205316}},
  \href{http://www-public.slac.stanford.edu/sciDoc/docMeta.aspx?slacPubNumber=SLAC-PUB-9228}{http://www-public.slac.stanford.edu/sciDoc/docMeta.aspx?slacPubNumber=SLAC-PUB-9228}.

\bibitem{Kachru:2003aw}
S.~Kachru, R.~Kallosh, A.~D. Linde and S.~P. Trivedi, \emph{{De Sitter vacua in
  string theory}},
  \href{https://doi.org/10.1103/PhysRevD.68.046005}{\emph{Phys. Rev.}
  {\bfseries D68} (2003) 046005}
  [\href{https://arxiv.org/abs/hep-th/0301240}{{\ttfamily hep-th/0301240}}].

\bibitem{Cicoli:2018kdo}
M.~Cicoli, S.~De~Alwis, A.~Maharana, F.~Muia and F.~Quevedo, \emph{{De Sitter
  vs Quintessence in String Theory}},
  \href{https://arxiv.org/abs/1808.08967}{{\ttfamily 1808.08967}}.

\bibitem{Akrami:2018ylq}
Y.~Akrami, R.~Kallosh, A.~Linde and V.~Vardanyan, \emph{{The landscape, the
  swampland and the era of precision cosmology}},
  \href{https://arxiv.org/abs/1808.09440}{{\ttfamily 1808.09440}}.

\bibitem{Kachru:2018aqn}
S.~Kachru and S.~P. Trivedi, \emph{{A comment on effective field theories of
  flux vacua}},  \href{https://arxiv.org/abs/1808.08971}{{\ttfamily
  1808.08971}}.

\bibitem{Hebecker:2018vxz}
A.~Hebecker and T.~Wrase, \emph{{The asymptotic dS Swampland Conjecture - a
  simplified derivation and a potential loophole}},
  \href{https://arxiv.org/abs/1810.08182}{{\ttfamily 1810.08182}}.

\bibitem{Conlon:2018eyr}
J.~P. Conlon, \emph{{The de Sitter swampland conjecture and supersymmetric AdS
  vacua}},  \href{https://arxiv.org/abs/1808.05040}{{\ttfamily 1808.05040}}.

\bibitem{Murayama:2018lie}
H.~Murayama, M.~Yamazaki and T.~T. Yanagida, \emph{{Do We Live in the
  Swampland?}},  \href{https://arxiv.org/abs/1809.00478}{{\ttfamily
  1809.00478}}.

\bibitem{Blaback:2018hdo}
J.~Blåbäck, U.~Danielsson and G.~Dibitetto, \emph{{A new light on the darkest
  corner of the landscape}},
  \href{https://arxiv.org/abs/1810.11365}{{\ttfamily 1810.11365}}.

\bibitem{Tsujikawa:2013fta}
S.~Tsujikawa, \emph{{Quintessence: A Review}},
  \href{https://doi.org/10.1088/0264-9381/30/21/214003}{\emph{Class. Quant.
  Grav.} {\bfseries 30} (2013) 214003}
  [\href{https://arxiv.org/abs/1304.1961}{{\ttfamily 1304.1961}}].

\bibitem{Agrawal:2018own}
P.~Agrawal, G.~Obied, P.~J. Steinhardt and C.~Vafa, \emph{{On the Cosmological
  Implications of the String Swampland}},
  \href{https://doi.org/10.1016/j.physletb.2018.07.040}{\emph{Phys. Lett.}
  {\bfseries B784} (2018) 271}
  [\href{https://arxiv.org/abs/1806.09718}{{\ttfamily 1806.09718}}].

\bibitem{Ashoorioon:2018sqb}
A.~Ashoorioon, \emph{{Rescuing Single Field Inflation from the Swampland}},
  \href{https://arxiv.org/abs/1810.04001}{{\ttfamily 1810.04001}}.

\bibitem{Motaharfar:2018zyb}
M.~Motaharfar, V.~Kamali and R.~O. Ramos, \emph{{Warm way out of the
  Swampland}},  \href{https://arxiv.org/abs/1810.02816}{{\ttfamily
  1810.02816}}.

\bibitem{Dimopoulos:2018upl}
K.~Dimopoulos, \emph{{Steep Eternal Inflation and the Swampland}},
  \href{https://arxiv.org/abs/1810.03438}{{\ttfamily 1810.03438}}.

\bibitem{Lin:2018kjm}
C.-M. Lin, K.-W. Ng and K.~Cheung, \emph{{Chaotic inflation on the brane and
  the Swampland Criteria}},  \href{https://arxiv.org/abs/1810.01644}{{\ttfamily
  1810.01644}}.

\bibitem{Das:2018hqy}
S.~Das, \emph{{A note on Single-field Inflation and the Swampland Criteria}},
  \href{https://arxiv.org/abs/1809.03962}{{\ttfamily 1809.03962}}.

\bibitem{Brahma:2018hrd}
S.~Brahma and M.~Wali~Hossain, \emph{{Avoiding the string swampland in
  single-field inflation: Excited initial states}},
  \href{https://arxiv.org/abs/1809.01277}{{\ttfamily 1809.01277}}.

\bibitem{Damian:2018tlf}
C.~Damian and O.~Loaiza-Brito, \emph{{Two-field axion inflation and the
  swampland constraint in the flux-scaling scenario}},
  \href{https://arxiv.org/abs/1808.03397}{{\ttfamily 1808.03397}}.

\bibitem{Matsui:2018bsy}
H.~Matsui and F.~Takahashi, \emph{{Eternal Inflation and Swampland
  Conjectures}},  \href{https://arxiv.org/abs/1807.11938}{{\ttfamily
  1807.11938}}.

\bibitem{Kehagias:2018uem}
A.~Kehagias and A.~Riotto, \emph{{A note on Inflation and the Swampland}},
  \href{https://arxiv.org/abs/1807.05445}{{\ttfamily 1807.05445}}.

\bibitem{Achucarro:2018vey}
A.~Achúcarro and G.~A. Palma, \emph{{The string swampland constraints require
  multi-field inflation}},  \href{https://arxiv.org/abs/1807.04390}{{\ttfamily
  1807.04390}}.

\bibitem{Lehners:2018vgi}
J.-L. Lehners, \emph{{Small-Field and Scale-Free: Inflation and Ekpyrosis at
  their Extremes}},  \href{https://arxiv.org/abs/1807.05240}{{\ttfamily
  1807.05240}}.

\bibitem{Lin:2018rnx}
C.-M. Lin, \emph{{Type I Hilltop Inflation and the Refined Swampland
  Criteria}},  \href{https://arxiv.org/abs/1810.11992}{{\ttfamily 1810.11992}}.

\bibitem{Park:2018fuj}
S.~C. Park, \emph{{Minimal gauge inflation and the refined Swampland
  conjecture}},  \href{https://arxiv.org/abs/1810.11279}{{\ttfamily
  1810.11279}}.

\bibitem{Ben-Dayan:2018mhe}
I.~Ben-Dayan, \emph{{Draining the Swampland}},
  \href{https://arxiv.org/abs/1808.01615}{{\ttfamily 1808.01615}}.

\bibitem{Chiang:2018jdg}
C.-I. Chiang and H.~Murayama, \emph{{Building Supergravity Quintessence
  Model}},  \href{https://arxiv.org/abs/1808.02279}{{\ttfamily 1808.02279}}.

\bibitem{Wang:2018kly}
S.-J. Wang, \emph{{Quintessential Starobinsky inflation and swampland
  criteria}},  \href{https://arxiv.org/abs/1810.06445}{{\ttfamily 1810.06445}}.

\bibitem{DAmico:2018mnx}
G.~D'Amico, N.~Kaloper and A.~Lawrence, \emph{{Strongly Coupled Quintessence}},
   \href{https://arxiv.org/abs/1809.05109}{{\ttfamily 1809.05109}}.

\bibitem{Han:2018yrk}
C.~Han, S.~Pi and M.~Sasaki, \emph{{Quintessence Saves Higgs Instability}},
  \href{https://arxiv.org/abs/1809.05507}{{\ttfamily 1809.05507}}.

\bibitem{Olguin-Tejo:2018pfq}
Y.~Olguin-Tejo, S.~L. Parameswaran, G.~Tasinato and I.~Zavala, \emph{{Runaway
  Quintessence, Out of the Swampland}},
  \href{https://arxiv.org/abs/1810.08634}{{\ttfamily 1810.08634}}.

\bibitem{Denef:2018etk}
F.~Denef, A.~Hebecker and T.~Wrase, \emph{{de Sitter swampland conjecture and
  the Higgs potential}},
  \href{https://doi.org/10.1103/PhysRevD.98.086004}{\emph{Phys. Rev.}
  {\bfseries D98} (2018) 086004}
  [\href{https://arxiv.org/abs/1807.06581}{{\ttfamily 1807.06581}}].

\bibitem{Hamaguchi:2018vtv}
K.~Hamaguchi, M.~Ibe and T.~Moroi, \emph{{The swampland conjecture and the
  Higgs expectation value}},
  \href{https://arxiv.org/abs/1810.02095}{{\ttfamily 1810.02095}}.

\bibitem{Choi:2018rze}
K.~Choi, D.~Chway and C.~S. Shin, \emph{{The dS swampland conjecture with the
  electroweak symmetry and QCD chiral symmetry breaking}},
  \href{https://arxiv.org/abs/1809.01475}{{\ttfamily 1809.01475}}.

\bibitem{Dvali:2018fqu}
G.~Dvali and C.~Gomez, \emph{{On Exclusion of Positive Cosmological Constant}},
   \href{https://arxiv.org/abs/1806.10877}{{\ttfamily 1806.10877}}.

\bibitem{Andriot:2018wzk}
D.~Andriot, \emph{{On the de Sitter swampland criterion}},
  \href{https://doi.org/10.1016/j.physletb.2018.09.022}{\emph{Phys. Lett.}
  {\bfseries B785} (2018) 570}
  [\href{https://arxiv.org/abs/1806.10999}{{\ttfamily 1806.10999}}].

\bibitem{Garg:2018reu}
S.~K. Garg and C.~Krishnan, \emph{{Bounds on Slow Roll and the de Sitter
  Swampland}},  \href{https://arxiv.org/abs/1807.05193}{{\ttfamily
  1807.05193}}.

\bibitem{Ooguri:2018wrx}
H.~Ooguri, E.~Palti, G.~Shiu and C.~Vafa, \emph{{Distance and de Sitter
  Conjectures on the Swampland}},
  \href{https://arxiv.org/abs/1810.05506}{{\ttfamily 1810.05506}}.

\bibitem{Bousso:1999xy}
R.~Bousso, \emph{{A Covariant entropy conjecture}},
  \href{https://doi.org/10.1088/1126-6708/1999/07/004}{\emph{JHEP} {\bfseries
  07} (1999) 004} [\href{https://arxiv.org/abs/hep-th/9905177}{{\ttfamily
  hep-th/9905177}}].

\bibitem{Roupec:2018mbn}
C.~Roupec and T.~Wrase, \emph{{de Sitter extrema and the swampland}},
  \href{https://arxiv.org/abs/1807.09538}{{\ttfamily 1807.09538}}.

\bibitem{Garg:2018zdg}
S.~K. Garg, C.~Krishnan and M.~Zaid, \emph{{Bounds on Slow Roll at the Boundary
  of the Landscape}},  \href{https://arxiv.org/abs/1810.09406}{{\ttfamily
  1810.09406}}.

\bibitem{Caviezel:2009tu}
C.~Caviezel, T.~Wrase and M.~Zagermann, \emph{{Moduli Stabilization and
  Cosmology of Type IIB on SU(2)-Structure Orientifolds}},
  \href{https://doi.org/10.1007/JHEP04(2010)011}{\emph{JHEP} {\bfseries 04}
  (2010) 011} [\href{https://arxiv.org/abs/0912.3287}{{\ttfamily 0912.3287}}].

\bibitem{Flauger:2008ad}
R.~Flauger, S.~Paban, D.~Robbins and T.~Wrase, \emph{{Searching for slow-roll
  moduli inflation in massive type IIA supergravity with metric fluxes}},
  \href{https://doi.org/10.1103/PhysRevD.79.086011}{\emph{Phys. Rev.}
  {\bfseries D79} (2009) 086011}
  [\href{https://arxiv.org/abs/0812.3886}{{\ttfamily 0812.3886}}].

\bibitem{Akrami:2018odb}
{\scshape Planck} collaboration, Y.~Akrami et~al., \emph{{Planck 2018 results.
  X. Constraints on inflation}},
  \href{https://arxiv.org/abs/1807.06211}{{\ttfamily 1807.06211}}.

\bibitem{Scolnic:2017caz}
D.~M. Scolnic et~al., \emph{{The Complete Light-curve Sample of
  Spectroscopically Confirmed SNe Ia from Pan-STARRS1 and Cosmological
  Constraints from the Combined Pantheon Sample}},
  \href{https://doi.org/10.3847/1538-4357/aab9bb}{\emph{Astrophys. J.}
  {\bfseries 859} (2018) 101}
  [\href{https://arxiv.org/abs/1710.00845}{{\ttfamily 1710.00845}}].

\bibitem{Ade:2015xua}
{\scshape Planck} collaboration, P.~A.~R. Ade et~al., \emph{{Planck 2015
  results. XIII. Cosmological parameters}},
  \href{https://doi.org/10.1051/0004-6361/201525830}{\emph{Astron. Astrophys.}
  {\bfseries 594} (2016) A13}
  [\href{https://arxiv.org/abs/1502.01589}{{\ttfamily 1502.01589}}].

\bibitem{Heisenberg:2018yae}
L.~Heisenberg, M.~Bartelmann, R.~Brandenberger and A.~Refregier, \emph{{Dark
  Energy in the Swampland}},
  \href{https://arxiv.org/abs/1808.02877}{{\ttfamily 1808.02877}}.

\bibitem{Heisenberg:2018rdu}
L.~Heisenberg, M.~Bartelmann, R.~Brandenberger and A.~Refregier, \emph{{Dark
  Energy in the Swampland II}},
  \href{https://arxiv.org/abs/1809.00154}{{\ttfamily 1809.00154}}.

\bibitem{Gott:2010xw}
J.~R. Gott and Z.~Slepian, \emph{{Dark Energy as Double
  N-Flation--Observational Predictions}},
  \href{https://doi.org/10.1111/j.1365-2966.2011.19049.x}{\emph{Mon. Not. Roy.
  Astron. Soc.} {\bfseries 416} (2011) 907}
  [\href{https://arxiv.org/abs/1011.2528}{{\ttfamily 1011.2528}}].

\bibitem{Slepian:2013ug}
Z.~Slepian, J.~R. Gott, III and J.~Zinn, \emph{{A one-parameter formula for
  testing slow-roll dark energy: observational prospects}},
  \href{https://doi.org/10.1093/mnras/stt2195}{\emph{Mon. Not. Roy. Astron.
  Soc.} {\bfseries 438} (2014) 1948}
  [\href{https://arxiv.org/abs/1301.4611}{{\ttfamily 1301.4611}}].

\bibitem{Fukuda:2018haz}
H.~Fukuda, R.~Saito, S.~Shirai and M.~Yamazaki, \emph{{Phenomenological
  Consequences of the Refined Swampland Conjecture}},
  \href{https://arxiv.org/abs/1810.06532}{{\ttfamily 1810.06532}}.

\bibitem{Das:2018rpg}
S.~Das, \emph{{Warm Inflation in the light of Swampland Criteria}},
  \href{https://arxiv.org/abs/1810.05038}{{\ttfamily 1810.05038}}.

\end{thebibliography}\endgroup

\end{document}